\documentclass[11pt,tightenlines,eqsecnum,floats,aps,amsmath,amssymb,nofootinbib,prd,shownopacs,floatfix]{revtex4}
\usepackage{graphicx, wrapfig}
\usepackage{amssymb}
\usepackage{color}
\usepackage{mathrsfs}
\usepackage{subfigure}
\def\f{\frac}
\def\dd{\textrm{d}}

\def\t{\tilde}
\def\h{\hat}
\def\b{\bar}
\def\vk{\vec k}
\def\vx{\vec{x}}
\def\qo{\mathring{q}}
\def\ps{\mathbf{\Gamma}}
\def\psE{\mathbf{\Gamma}_{\rm Ext}}

\def\lp{l_{\rm Pl}}

\def\rhopl{\rho_{\rm Pl}}
\def\rhosup{\rho_{_{\rm sup}}}

\def\Hp{\mathcal{H}_{\rm phy}}

\def\B{\mathcal{B}}

\def\hRone{\hat{\mathfrak{R}}}
\def\hBone{\hat{B}}
\def\Eone{{}^{(1)}\!E}
\def\Bone{{}^{(1)}\!B}
\def\Rone{{}^{(1)}\mathfrak R}

\def\Rthree{\mathfrak R}

\def\Dzero{\mathring D}

\def\ak{\hat{a}_{\vec{k}}}

\def\admk{\hat{a}_{-\vec{k}}^\dagger}

\usepackage{enumerate}

\newcommand{\ig}{\includegraphics}
\newcommand{\be}{\nopagebreak[3]\begin{equation}}
\newcommand{\ee}{\end{equation}}
\newcommand{\bfig}{\nopagebreak[3]\begin{figure}}
\newcommand{\efig}{\end{figure}}
\newcommand{\ba}{\nopagebreak[3]\begin{eqnarray}}
\newcommand{\ea}{\end{eqnarray}}

\newcommand{\bmult}{\nopagebreak[3]\begin{multline}}
\newcommand{\emult}{\end{multline}}
\newcommand{\fref}[1]{Fig.\,\ref{#1}}

\begin{document}
\title{Initial conditions for cosmological perturbations}\author{Abhay Ashtekar${}^{1,2}\,$}
\email{ashtekar@gravity.psu.edu} 
\author{Brajesh Gupt${}^{1}\,$}
\email{bgupt@gravity.psu.edu}
\affiliation{
${}^{1}$Institute for Gravitation and the Cosmos \& Physics Department, The Pennsylvania State University, University Park, PA 16802 U.S.A.\\
\\
${}^{2}$Perimeter Institute for Theoretical Physics, 31 Caroline St N, Waterloo, Ontario, Canada N2L 2Y5
}


\begin{abstract}
Penrose proposed that the big bang singularity should be constrained by requiring that the Weyl curvature vanishes there. The idea behind this past hypothesis is attractive because it constrains the initial conditions for the universe in geometric terms and is not confined to a specific early universe paradigm. However, the precise statement of Penrose's hypothesis is tied to classical space-times and furthermore restricts only the gravitational degrees of freedom. These are encapsulated only in the tensor modes of the commonly used cosmological perturbation theory. Drawing inspiration from the underlying idea, we propose a quantum generalization of Penrose's hypothesis using the Planck regime in place of the big bang, and simultaneously incorporating tensor as well as scalar modes. Initial conditions selected by this generalization constrain the universe to be as homogeneous and isotropic in the Planck regime \emph{as permitted by the Heisenberg uncertainty relations}.
\end{abstract}

\maketitle

\section{Introduction}
\label{s1}

The issue of initial conditions for the universe has been debated extensively in the literature. Some find it natural to assume that the initial conditions were \emph{generic}, and the large scale homogeneity and isotropy we observe today arose from dynamics. In particular, inflation is often invoked as the key mechanism behind this phenomenon. For example, in the 1980s it was suggested that space-time was irregular at all scales during the Planck era, representing a thermal foam of maximum entropy. But, because inflation is efficient in diluting irregularities, already by the end of inflation the universe reached the uniformity we observe on large scale today (see, e.g., \cite{davies} and references therein). However, to get inflation \emph{started} within the scenarios that were then contemplated, the universe would have had to cool below the GUT scale from the thermal foam in the Planck epoch, and it was unclear as to why and how this should occur \cite{vilenkin}. These lines of reasoning suggested the opposite viewpoint that the initial state had to be \emph{very special}, e.g., without long range spatial correlations in fluctuations of geometry and matter (see, e.g., \cite{page}). A specific proposal in this direction is Penrose's Weyl curvature hypothesis (WCH) which posits that, in spite of the strong curvature  singularity, big bang is very special in that the Weyl curvature vanishes there \cite{rp,rp1}. The proposal is attractive because it is completely general and not tied to the details of any specific scenario used to describe the early universe. 

We share the underlying viewpoint that a past hypothesis is needed to significantly narrow down initial conditions and account for the extraordinary homogeneity and isotropy of the early universe. However, we believe that one needs to extend the original WCH in two directions. First, it should not be formulated in the context of classical gravity, but take into account important quantum effects. Indeed we know that, already at nuclear densities, quantum physics is crucially important in astronomy to account for the very existence of neutron stars. Therefore, there is every expectation that quantum effects would become dominant in the Planck epoch when densities are some $10^{80}$ times the nuclear density. Thus, we need an extension of the WCH that refers to the Planck regime of quantum gravity rather than the big bang of classical general relativity (GR). Secondly, while Weyl curvature does vanish in spatially homogeneous and isotropic space-times, the condition that it should vanish does not by itself imply that the space-time metric $g_{ab}$ must be spatially homogeneous and isotropic since the conformal factor relating $g_{ab}$ to the flat metric can have arbitrary space-time dependence. Therefore restrictions on the Weyl tensor do not by themselves suffice to arrive at the initial homogeneity and isotropy. The goal of this paper is to generalize the WCH in order to overcome these two limitations. 

Our understanding of the early universe has evolved significantly over the past three decades. Some of the ideas that featured prominently in the 1980s have been transcended or even ruled out, and a mainstream approach to investigate the primordial universe has emerged. In this approach the early universe is described by a Friedmann, Lema\^{i}tre, Robertson, Walker (FLRW) background space-time, together with cosmological perturbations described by \emph{quantum} fields. While this is a much more restricted paradigm compared to what was contemplated in the semi-qualitative discussions in the 1980s, we still face the issue of initial conditions since there is considerable freedom in the choice of the FLRW background as well as the quantum state of perturbations. Furthermore, since the task is now defined sharply, it calls for a treatment that is mathematically precise and more detailed than before. 

If one works with a classical FLRW background, the problem becomes ill-posed because one would have to impose initial conditions on cosmological perturbations at the big bang singularity. Fortunately, this obstacle can now be overcome because our understanding of the \emph{quantum} FLRW geometry in the Planck regime has also evolved significantly. In particular, there is a rich body of literature in Loop Quantum Cosmology (LQC) which shows that the big bang singularity is naturally resolved by quantum geometry effects (for reviews, see, e.g., \cite{asrev,ps3,agullocorichi}). In this framework, there is a Hilbert space $\Hp$ of physical quantum states, consisting of solutions to the quantum Hamiltonian constraint, and operators representing observables such as the matter density, curvature, anisotropies, etc. on $\Hp$. Together, they provide a specific and detailed description of the quantum space-time geometry in the Planck regime.  A surprising aspect of this framework is that --much like the early universe-- the Planck regime appears to be tamer than what one would have a priori imagined; there are no thermal foams or fractals that were envisaged in the early discussions. On the other hand, fundamental discreteness underlying this quantum geometry \cite{alrev,crbook,ttbook} is  much more subtle than what was commonly assumed, resulting in  unforeseen interplays between the ultraviolet and the infrared (see, e.g. \cite{aan3,agulloassym,ag3}). For concreteness, we will work in the general paradigm provided by LQC. However, since our analysis uses only the qualitative features of quantum geometry near the bounce, the generalized WCH can be used also in other bouncing scenarios.
  
Thus, we will assume that the big bang singularity is resolved by quantum gravity effects and there is a bounce in a regime in which the space-time curvature and matter density are of Planck scale. For concreteness, we will assume that the background is described by a spatially flat \emph{quantum} FLRW geometry. The generalization of the WCH will be formulated for the system consisting of this background quantum geometry, \emph{together with} quantum fields representing cosmological perturbations on this background. 

Let us first consider tensor modes. They give rise to a Weyl tensor \emph{operator} $\h{C}_{abc}{}^{d}$ which is subject to non-trivial commutation relations. Consequently, one cannot carry over Penrose's formulation directly. To understand the conceptual obstacle, let us first consider the quantum theory of the Maxwell field in Minkowski space-time. There is \emph{no} non-zero vector in the photon Fock space on which the Maxwell field operator $\h{F}_{ab}$ vanishes identically: the commutator between its  electric and magnetic fields $\h{E}_{a},\, \h{B}_{a}$ is a c-number whence, if a state were to be annihilated by $\h{F}_{ab}$, we would have a conflict with the Heisenberg uncertainty relations. For a completely analogous reason, there is no physical state of the system under consideration on which $\h{C}_{abc}{}^{d}$ vanishes identically. Thus, the generalized WCH has to respect the \emph{operator character} of the Weyl tensor --especially the Heisenberg uncertainty relations it is subject to. We will see that this requirement leads one to refine the WCH in several directions. Finally, Weyl tensor considerations are not appropriate to restrict the initial conditions for scalar modes, e.g., because these modes can not give rise to a non-trivial magnetic part of Weyl curvature. Our formulation of the generalized WCH for tensor modes will suggest a natural extension that is appropriate for scalar modes. In the end, our proposal will 
restrict the initial conditions on tensor and scalar modes to ensure maximal homogeneity and isotropy in the Planck regime, permissible within Heisenberg uncertainties. Therefore we will refer to the proposal as a \emph{quantum homogeneity and isotropy hypothesis} (QHIH).

The accompanying paper \cite{ag3} introduces another principle to severely restrict the background quantum geometry. It then uses the results of this paper to work out the phenomenological consequences of the choice of initial conditions for the total system within an LQC extension of the inflationary paradigm. The main result is that the ensuing power spectrum for scalar modes is in better agreement with observations of the PLANCK mission than standard inflation: while there is agreement with the standard predictions at small angular scales, thanks to the interplay between the ultraviolet and the infrared referred to above,  there is an appropriate power suppression at large angular scales. Thus, there is an interesting relation between initial conditions chosen in the Planck epoch and the CMB observations that refer to a time some 380,000 years later. We mention these results only to put the findings of the present paper in a broader perspective. The present paper does not assume inflation nor, as already mentioned, are the details of LQC dynamics necessary to the quantum generalization of the WCH. Therefore our QHIH could be useful well beyond the application discussed in \cite{ag3}. 

The paper is organized as follows. In section \ref{s2} we present the underlying framework. Specifically, the first part of this section considers tensor modes, introduces expressions of the electric and magnetic parts of the Weyl tensor in terms of metric perturbations and specifies the fundamental Poisson brackets. Phase space considerations will already lead to the first refinement of the WCH. The second part summarizes the salient features of the quantum FLRW geometry in the Planck regime. Our QHIH is presented in section \ref{s3}. After discussing the main idea, we implement it first in the simpler setting of flat space-time in section \ref{s3.1}. This mathematical detour will enable us to streamline calculations and also bring out the conceptual complications one encounters in the quantum FLRW geometry. These are addressed in section \ref{s3.2}. In the classical theory, to ensure homogeneity and isotropy it is necessary and sufficient that certain fundamental observables vanish on the canonical phase space. The final formulation of our QHIH requires that, in the Planck regime, expectation values of those observables should vanish and their uncertainties should be as small as they are allowed to be by the uncertainty relations. This condition selects a ball in the space of all quasi-free Heisenberg states \cite{quasi-free} of tensor modes, thereby severely restricting the initial conditions during the Planck epoch. For scalar modes, the construction is completely analogous and is summarized at the end of each subsection. In section \ref{s4} we summarize the main results and put them in a broader context. 

Our conventions are as follows. We use signature -,+,+,+ and set $c=1$.  But keep $G$ and $\hbar$ explicitly in various equations to facilitate the distinction between classical and quantum effects. As is usual, we will set $\kappa = 8\pi G$. To  bypass certain infrared complications which are irrelevant for the physics under discussion, we will assume that the spatial topology is that of a 3-torus $\mathbb{T}^{3}$.
Since flat metrics on $\mathbb{T}^{3}$ admit only 3 global Killing fields (which are `translational'), there metrics are globally homogeneous and only locally isotropic. Therefore, by `isotropy' we will mean only local isotropy. 
It is straightforward but somewhat cumbersome to extend our entire construction from $\mathbb{T}^{3}$ topology to $\mathbb{R}^{3}$ by appropriately handling the distributional character of the Fourier modes of quantum fields representing cosmological perturbations.

\section{Underlying framework}
\label{s2}

In this section we introduce the basic concepts and mathematical relations that will be used throughout the rest of the paper. 
In section \ref{s2.1} we discuss tensor perturbations on FLRW backgrounds and express the electric and magnetic parts of the Weyl tensor in terms of the two tensor modes that are commonly used. Since the QHIH is based on Heisenberg uncertainties, we present the Poisson bracket relations between various geometric observables that characterize departure from homogeneity and isotropy. This classical framework will provide the point of departure in section \ref{s3}. 
In section \ref{s2.2} we summarize the key properties of quantum FLRW space-times and introduce a convenient characterization of what is meant by `the Planck regime'. 

\subsection{Tensor modes: Electric and magnetic parts of the Weyl tensor}
\label{s2.1}

Consider a metric $\b{g}_{ab}$ representing a perturbed FLRW metric $\mathring{\b{g}}_{ab}$,
\be \label{gbar1}
  \b{g}_{ab} = \mathring{\b{g}}_{ab} +\epsilon \b{h}_{ab},
\ee
where $\epsilon$ is a smallness parameter and $\b{h}_{ab}$ denotes the first order perturbation. The background FLRW metric has the form
\be \mathring{\bar g}_{ab} \dd x^a \dd x^b = -\dd t^2 + a^2 \dd {\vx}^2  
= a^{2} (-\dd\eta^{2} + \dd {\vx}^2) \ee
where, as usual, $t$ denotes the proper time, $\eta$ the conformal time and $a$ is the scale factor. In the main body of the paper, we restrict ourselves to tensor modes. Therefore, in the Lorentz and radiation gauge,
\be \label{gauge1} \b{\nabla}^{a} \b{h}_{ab}=0,\quad \b{h}_{ab}\eta^{b} =0,\quad {\rm and} \quad \b{h}_{ab}\b{g}^{ab} =0\, , \ee
$h_{ab}$ satisfies 
\be \label{fe1}
\b{\Box}\, \b{h}_{ab} - 2H^{2}\, \bar{h}_{ab} =0\, . \ee
Here $\eta^{a}\partial_{a} = \partial/\partial \eta$ is normal to the cosmological slices and $H= \dot{a}/a$ is the Hubble rate. As is well known, the analysis simplifies considerably if one rewrites (\ref{gbar1}) as 
\be \label{gbar2}
  \b{g}_{ab} = a^2(\eta) g_{ab}  
              = a^2(\eta) \left(\mathring{g}_{ab} + \epsilon h_{ab}\right),
\ee
and works with $h_{ab}$ in place of $\b{h}_{ab}$ and the flat metric $\mathring{g}_{ab}$ in place of the FLRW metric $\mathring{\b{g}}_{ab}$. For, the gauge conditions (\ref{gauge1}) can now be written using the \emph{flat} metric $\mathring{g}_{ab}$ 
\be \label{gauge2}
\mathring{\nabla}^{a} h_{ab} =0, \quad h_{ab}\eta^{a}=0, \quad {\rm and} \quad h_{ab}\mathring{g}^{ab} =0\, ,\ee
and the field equation (\ref{fe1}) reduces to 
\be \label{fe2}
  \mathring{\Box} h_{ab} - 2\f{a^{\prime}}{a}\, h^{\prime}_{ab} \equiv
  - h^{\prime\prime}_{ab} + \Dzero^2 h_{ab} - 2\f{a^{\prime}}{a}\, h^\prime_{ab} = 0\, ,
\ee
where the \emph{prime denotes derivative w.r.t.} $\eta$. Thus, except for the term $2 \f{a^{\prime}}{a}\, h^{\prime}_{ab}$ in (\ref{fe2}), the field $h_{ab}$ satisfies the \emph{same gauge conditions and dynamical equation as the tensor perturbation does on the flat background} $\mathring{g}_{ab}$. We will return to this fact in section \ref{s3.1}. It is convenient to regard  $\mathring{g}_{ab}$ as background metric and $h_{ab}$ a perturbation propagating on this flat background. We will adopt this viewpoint. In particular, all indices will be raised and lowered using $\mathring{g}_{ab}$.

Because the metrics $\b{g}_{ab}$ and $g_{ab}$ are conformally
related, their Weyl tensors satisfy $\b{C}_{abc}{}^{d} = C_{abc}{}^{d}$, whence the electric and magnetic parts of the two Weyl tensor are equal
\be \b{E}_{ab} = E_{ab} := {C}_{amb}{}^{n}\, \eta^{m} \eta_{n},\quad {\rm and} \quad \b{B}_{ab} = B_{ab} := {}^{\star}\!{C}_{amb}{}^{n}\, \eta^{m} \eta_{n} \ee
since $\eta^{m}$ is the unit normal to cosmological slices w.r.t. $\mathring{g}_{ab}$. Now, given \emph{any} 4-metric $g_{ab}$, one can express its $E_{ab}$ and $B_{ab}$ on a 3-slice using the 4-dimensional Ricci tensor $R_{ab}$ and the initial data --the intrinsic metric $q_{ab}$ and the extrinsic curvature $K_{ab}$ on the slice via  
\ba
  E_{ab} &=& \Rthree_{ab} - K_a{}^mK_{bm} + K K_{ab} -\f{1}{2}
            \left(q_a{}^mq_b{}^n+q_{ab}q^{mn}\right) 
            \left(R_{mn}-\f{1}{6} R~g_{mn}\right) \label{E1}\\
   B_{ab} &=& \epsilon_{mn(a}\, D^mK^n{}_{b)}\, , \label{B1}
\ea
where $D_{a},\,\Rthree_{ab}$ and $\epsilon_{abc}$ are the derivative operator, the Ricci tensor, and the volume 3-form of the spatial metric $q_{ab}$. Using the fact that our metric $g_{ab}$ is given by $g_{ab} = \mathring{g}_{ab} + \epsilon\, h_{ab}$, and $\mathring{g}_{ab}$ is flat, we find that the linearized Ricci tensor ${}^{(1)}{\Rthree}_{ab}$ and extrinsic curvature ${}^{(1)}\!K_{ab}$ are given by 
\be  {}^{(1)}{\Rthree}_{ab} = -\f{1}{2}\, \mathring{D}^{2} h_{ab} \quad {\rm and} \quad {}^{(1)}\!K_{ab} = \f{1}{2}\, h^{\prime}_{ab}\, ,  \ee
where $\mathring{D}$ is the derivative operator defined by the flat 3-metric $\mathring{q}_{ab}$ on the $\mathbb{T}^{3}$ spatial sections,
tailored to the co-moving coordinates ${\vx}$. Next, since $\mathring{g}_{ab}$ is flat, the 4-Ricci tensor $R_{ab}$ and the extrinsic curvature $K_{ab}$ vanish in the background. Therefore, using (\ref{fe2}) the linearized electric and magnetic parts $\Eone_{ab},\, \Bone_{ab}$ can be expressed in terms of the tensor modes $h_{ab}$ as
\ba
  \Eone_{ab} &=& 
{}^{(1)}{\Rthree}_{ab} + \f{1}{4} \mathring{\Box}\, h_{ab} 
             \label{E2}\\
 \Bone_{ab} &=& ~\mathring\epsilon_{(a}{}^{mn} \Dzero_{|m|} {}^{(1)}\!K_{b)n} . \label{B2}
\ea

Next, we note that the two radiative modes of the tensor perturbations can be extracted most directly by performing a spatial Fourier transform%
\be
 h_{ab}(\eta, {\vx}) = \f{1}{V_o}\, \sum_{s=1}^{2}\,\,\sum_{\vk} \,\,h_{\vec{k}}^{(s)}(\eta)\,\,e_{ab}^{(s)}(\vec{k})\,\,e^{i{\vk}\cdot{\vx}}\, 
\ee
where $V_{o}$ denotes the volume of the spatial $\mathbb{T}^{3}$ defined by $\mathring{q}_{ab}$,\,\, $(s)$ labels the two helicity states, and $e_{ab}^{(s)}$ are the polarization tensors satisfying
\ba
 e_{[ab]}^{(s)}({\vk}) &=& 0; \qquad k^a e_{ab}^{(s)}({\vk}) = 0;  \qquad \mathring q^{ab} e_{ab}^{(s)}({\vk}) = 0 \nonumber \\
 \left( e_{ab}^{(s)}({\vk})\right)^\star &=&  e_{ab}^{(s)}(-{\vk}); \qquad e_{ab}^{(s)}({\vk}) \, e_{cd}^{(s^\prime)}(-{\vk}) \,\mathring q^{ca} \,\mathring q^{db}= \delta_{s,s^\prime} \, .
\label{poltensor}
\ea
Here and in what follows $\star$ denotes the complex conjugate. Thus the two tensor modes of gravitational waves are captured in the gauge invariant functions $h_{\vk}^{(s)}$. Properties (\ref{poltensor}) ensure that the gauge conditions (\ref{gauge2}) are satisfied and the dynamical equation (\ref{fe2}) reduces to 
\be \label{fe3}
(h^{(s)}_{\vk})^{\prime\prime}+2\f{a^\prime}{a} (h^{(s)}_{\vk})^{\prime}+k^2 h^{(s)}_{\vec{k}}=0\, .
\ee

The phase space $\ps$ of tensor perturbations can now be specified as follows (see, e.g., \cite{aan2}). Consider an abstract 3-manifold $M$ which is topologically $\mathbb{T}^{3}$ and fix a fiducial, positive definite flat metric $\qo_{ab}$ on it. The basic variables are the radiative modes $h_{ab}^{(s)}$ and their conjugate momenta. It is simplest to work in the momentum space and rescale $h_{\vk}^{(s)}$ using $\kappa = 8\pi G$ to obtain configuration variables 
\be \label{rescaling} \phi_{\vk}^{(s)} := \f{1}{\sqrt{4\kappa}}\, h_{\vk} \ee
with physical dimensions of a scalar field (so that in section \ref{s3}  we will be able to use creation and annihilation operators with the standard normalization from the theory of scalar fields in FLRW space-times \cite{aan2}). The $\phi_{\vk}^{(s)}$ and their conjugate momenta $\pi_{\vk}^{(s)}$ satisfy the `reality conditions' 
\be \big(\phi_{\vk}^{(s)}\big)^{\star} = \phi_{-\vk}^{(s)} \qquad {\rm and}\qquad \big(\pi_{\vk}^{(s)}\big)^{\star} = \pi_{-\vk}^{(s)}\, ,\ee
and the only non-vanishing Poisson brackets between them are
\be \{\phi_{\vk}^{(s)}, \,\, \pi_{\vk^{\prime}}^{(s^{\prime})}\} =  V_{o}\,\,\delta_{\vk,\, -\vk^{\prime}}\,\, \delta^{s,s^{\prime}} 
\label{pb1} \ee
where $V_{o}$ is now the volume of $M$ with respect to $\qo_{ab}$. The Hamiltonian is time dependent, given by 
\be H(h, \pi;\, \eta) = \f{1}{2\,V_{o}}\, \sum_{s=1}^{2}\,\,\sum_{\vk}\,\, k^{2} a^{2}(\eta)|\phi_{\vk}^{(s)}|^{2}\, +\, a(\eta)^{-2}|\pi_{\vk}^{(s)}|^{2} \, .  \ee
It is easy to verify that the resulting equations of motion on $\ps$ reproduce (\ref{fe3}) in space-time. As with any system with a time dependent Hamiltonian, it is appropriate to work with an extended phase space $\psE = \ps \times \mathbb{R}$ where the $\mathbb{R}$ direction is coordinatized by time $\eta$. Then each leaf $\ps_{\eta}$ of $\psE$ carries the memory of the scale factor $a(\eta)$, the Hamiltonian is simply a function on $\psE$ and dynamics is represented by a genuine flow on $\psE$ \cite{aa-unitarity}.

Since we are interested in the Heisenberg uncertainties between various curvature quantities, let us examine corresponding observables (i.e. functions) on $\psE$ and their Poisson brackets. The intrinsic and extrinsic curvature at any time $\eta$ are completely captured in the observables
\ba \Rthree^{(s)}_{\vk}(\eta) &:=& \int \dd^{3}{x}\,\, {}^{(1)}\Rthree^{ab}(\vx, \eta)\, e_{ab}^{(s)}(\vk)\, e^{-i{\vk}\cdot{\vx}} \,\equiv\,  \sqrt{\kappa}\, k^{2}\, \phi_{\vk}^{(s)}(\eta) \qquad {\rm and} \label{Rthree}\\
K^{(s)}_{\vk}(\eta) &:=& \int \dd^{3}{x}\,\, {}^{(1)}\!K^{ab}(\vx, \eta)\, e_{ab}^{(s)}(\vk)\, e^{-i{\vk}\cdot{\vx}} \,\equiv\, \f{\sqrt{\kappa}}{a^{2}(\eta)}\, \pi_{\vk}^{(s)}(\eta) \label{K} \,\, \ea
and the only non-vanishing Poisson brackets between them are:
\be \{\Rthree_{\vk}^{(s)},\,\, K_{\vk^{\prime}}^{(s^{\prime})}\}
    = \kappa V_{o}\, \f{k^{2}}{a^{2}} \,\, \delta_{\vk,\, -\vk^{\prime}}\,\, \delta^{s,s^{\prime}} \ee

To formulate a quantum generalization of the WCH, let us return to the electric and magnetic parts, $\Eone_{ab}$ and $\Bone_{ab}$. The first obstruction to taking over the WCH directly to the quantum regime comes from the fact that \emph{$\Eone_{ab}$ is not a phase space observable} because it cannot be expressed as a function on $\psE$: the second term on the right hand side of (\ref{E2}) involves \emph{second} time-derivatives $h_{ab}(\eta, \vx)$. Therefore in the quantum generalization of the WCH, we will need to replace $\Eone_{ab}$ with another field. On the other hand, it is clear from (\ref{B2}) that \emph{$\Bone_{ab}$ is a phase space observable.} It is encoded in phase space functions 
\be B^{(s)}_{\vk}(\eta) := \int \dd^{3}{x}\,\, \Bone^{ab}(\vx, \eta)\,\, e^{(s)}_{m(a} (\vk)\,\, \epsilon_{b)}{}^{mn}(\vk)\,\,i\,\check{k}_{n}\,\, e^{-i{\vk}\cdot{\vx}} \,\equiv\, \f{\sqrt{\kappa}\, k}{a^{2}(\eta)} \pi_{\vk}^{(s)}(\eta) \, . \label{B3} \ee
(Note that while in (\ref{Rthree}) and (\ref{K}) smearing of $\Rone_{ab}$ and ${}^{(1)}\!K_{ab}$ is done using proper tensor fields, $\Bone_{ab}$ is smeared using pseudo tensor fields in (\ref{B3}) because $\Bone_{ab}$ is itself a pseudo tensor. Therefore like $\Rthree_{\vk}^{(s)}$ and $K^{(s)}_{\vk}$, the $B^{(s)}_{\vk}$ are also proper observables with respect to spatial reflections.) The $B_{\vk}^{(s)}$ satisfy the same reality condition 
\be \big(B^{(s)}_{\vk}\big)^{\star} = B^{(s)}_{-\vk}\, ,\ee
as $\Rthree^{(s)}_{\vk}$; have the same physical dimensions as $\Rthree^{(s)}_{\vk}$; and can be regarded as being `canonically conjugate' to them since
\be \label{RBPB} \{ \Rthree_{\vk}^{(s)}(\eta),\,\, B_{\vk^{\prime}}^{(s^{\prime})}(\eta) \}
    =   \f{\kappa}{a^{2}(\eta)}\, k^{3} V_{o}\, \delta_{\vk,\,-\vk^{\prime}}\,\, \delta_{s,s^{\prime}}\, . \ee 
Finally, note that ${}^{(1)}\Rthree_{ab}$ is gauge invariant and, as Eq (\ref{E2}) shows, it is the part of $\Eone_{ab}$ that does not refer to equations of motion. Tensor modes of cosmological perturbations vanish --i.e. spatial homogeneity and isotropy is maintained to first order-- if and only if all the phase space observables $\Rone^{(s)}_{\vk}$ and $\Bone^{(s)}_{\vk}$ (or, equivalently, $K^{(s)}_{\vk}$) vanish. Therefore, it is natural to use $\Rone_{ab}$ in place of $\Eone_{ab}$ in a formulation of the WCH on the canonical phase space, and hence in a  quantum generalization of the WCH. We will formulate our QHIH using expectation values and uncertainties in these observables, evaluated in the Planck regime.

\emph{Since the two helicities are decoupled, for simplicity of notation from now on we will drop the helicity index (s).} Thus symbols $\Rthree_{\vk}$ and $B_{\vk}$ will refer to observables corresponding to either one of the two helicities. \\

\emph{Remark}: $\Eone_{ab}$ differs from $\Rone_{ab}$ because of the term $\mathring{\Box} h_{ab}$ (see Eq. (\ref{E2})) and it is this term with second time derivatives that prevents $\Eone_{ab}$ from being an observable on the extended canonical phase space $\psE$. The situation in the linear theory simply mimics that in full general relativity where $E_{ab}$ \emph{cannot} be expressed as a function of phase space variables --or, of initial data $(q_{ab}, K_{ab})$ alone-- because of the presence of the 4-dimensional Ricci tensor in (\ref{E1}). Similarly, the fact that $\Bone_{ab}$ \emph{can be} expressed as a function on $\psE$ simply mimics the fact that in full general relativity $B_{ab}$ \emph{is} determined by the initial data (see Eq. (\ref{B1})).

When the linearized EOM are satisfied, the term $\Box h_{ab}$ can be replaced by\, $ 2 (a^{\prime}/a)\, h^{\prime}_{ab}$ (see Eq. (\ref{fe3})). Therefore, one can imagine constructing phase space functions using\! ${}^{(1)}\Rthree_{ab} + 2\sqrt{4\kappa} (a^{\prime}/a^{3})\, \pi_{ab}$ in place of $\Eone_{ab}$. But on the phase space, physical or geometrical significance of these functions would be unclear. Furthermore, it is easy to verify that the Poisson brackets between these phase space functions and $B^{(s)}_{\vk}$  are the same as those between $\Rthree^{(s)}_{\vk}$ and $B^{(s)}_{\vk}$. Therefore in considerations involving uncertainty, it is best to avoid an ad hoc use of field equations  and just use $\Rone_{ab}$ in place of $\Eone_{ab}$.

\subsection{The Planck regime}
\label{s2.2}

To obtain a quantum generalization of the WCH, we need to work not with the big bang singularity of classical GR, but with the Planck regime of a suitable quantum gravity theory. As explained in section \ref{s1}, the LQC bounce provides motivation for certain aspects of our proposal. Therefore, we will now sketch the principal features of this paradigm. Our proposal does not depend on the specific numbers that will feature in this summary. They are only meant to provide orders of magnitude to bring out the fact that there should be no difficulty in implementing the proposal in detail within specific scenarios of the early universe. (Indeed, it was implemented in the inflationary paradigm in \cite{ag3}.)

In LQC, a large number of cosmological models have been analyzed in detail \cite{asrev,ps3,agullocorichi}. In all cases, strong curvature singularities are naturally resolved. The mechanism can be traced back to the specific quantum Riemannian geometry that underlies Loop Quantum Gravity (LQG) (for summaries see, e.g., \cite{alrev,crbook,ttbook}). In LQG, the basic operators that generate the Heisenberg algebra are not the 3-metric and its conjugate momentum as in geometrodynamics, but holonomies $\h{h}_{\gamma}$ of the gravitational spin-connection along curves $\gamma$ and fluxes $\h{F}_{S}$ of orthonormal triads across 2-surfaces.%
\footnote{There is a precise sense in which $\h{F}_{S}$ are also the analogs of the fluxes of Yang-Mills electric field across surfaces $S$.} 
Geometric operators such as areas of 2-surfaces are defined using triad fluxes $\h{F}_{S}$. Riemannian geometry is quantized in the precise sense that all eigenvalues of geometric operators are discrete. In particular, there is a lowest non-zero eigenvalue of the area operator, denoted $\Delta$ (in Planck units) \cite{al5,rs,alarea}. It is called the \emph{area gap} and plays an important role in the Planck regime.  In standard Riemannian geometry, curvature of the spin-connection can be expressed as the limit of the ratio $h_{\gamma}/{\rm area_{\gamma}}$ of the holonomy around a closed loop $\gamma$ and the area enclosed by the loop, as the area tends to zero. In LQC, the analogous procedure has to respect the fundamental discreteness of the underlying quantum geometry. Hence the curvature operator is realized as the holonomy around the loop \emph{when the area it encloses shrinks to the area gap}, $\Delta$. As a result, $\Delta$ plays an important role in the dynamics generated by the quantum Hamiltonian constraint.  Specifically, in cosmological models, quantum geometry corrections create an effective `repulsive force' whose origin can be traced to $\Delta$. During most of the history of the universe, this force is completely negligible. However, it grows \emph{very} rapidly in the Planck regime, overwhelms the classical gravitational attraction and causes the universe to bounce, even when all matter fields satisfy the dominant energy conditions. 

For the purpose of this paper, it suffices to restrict ourselves to the spatially flat, $\Lambda=0$ quantum FLRW geometries \cite{bojowaldprl01,aps3,ps09} (although incorporation of spatial curvature \cite{apsv,warsaw1} and a non-zero cosmological constant \cite{bp,pa} would be completely analogous). For concreteness, suppose the only matter is a scalar field $\phi$. Then, there is a well-defined physical Hilbert space $\Hp$ consisting of wave functions $\Psi_{o}(v,\phi)$ satisfying the quantum Hamiltonian constraint where $v$ denotes the volume of the spatial hypersurfaces $\mathbb{T}^{3}$. The scalar field $\phi$ is generally used as a relational time variable. A complete set of Dirac observables is provided by the true Hamiltonian generating evolution with respect to $\phi$, and the operators $\hat{V}|_{\phi}$ that measure the spatial volume at time $\phi$ (or, the matter density operators $\h{\rho}|_{\phi}$). One can show that the universe bounces in the sense that, for \emph{every} physical state, the expectation values of the volume operators $\h{V}|_{\phi}$ have a non-zero minimum and those of the energy density operators $\h{\rho}|_{\phi}$  have a finite upper bound. Furthermore, the energy density operator has an absolute upper bound $\rhosup$ on the entire $\Hp$, given by
\be \rhosup\, = \,  \frac{{\rm 18\pi}}{G^{2}\hbar\, \Delta^{3}}\,\, \approx \,\,0.41\, \rhopl\, \ee
Note that in the limit in which the area gap $\Delta$ goes to zero --i.e. in the limit in which one ignores quantum geometry effects-- we obtain the classical GR result that the energy density has no upper bound and can diverge, giving rise to a strong curvature singularity.

If one asks that the physical state $\Psi_{o}$ be sharply peaked around a classical trajectory \emph{at late times}, one finds that they remain sharply peaked on an effective trajectory \emph{even in the Planck regime}. However, this trajectory undergoes a bounce because the `repulsive force' dominates in the Planck regime. We will restrict ourselves to such states because we know that the universe is extremely well described by GR at late times. For these $\Psi_{o}$ the space-time metric $\underbar{g}_{ab}$ defined by the effective trajectory is smooth everywhere. But of course it deviates very significantly from the GR trajectory in the Planck regime. At the bounce, the matter density attains its supremum $\rhosup \approx 0.41\, \rhopl$ and the scalar curvature reaches its maximum value, $R_{\rm sup}\approx 62\, \lp^{-2}$. This is just as one might expect; quantities that diverge at the big bang in GR remain finite and attain Planck scale values at the bounce. However, the situation is very different for the Hubble rate: While it too diverges at the big bang in GR, it \emph{vanishes} at the bounce (as it must because the universe is in a contracting phase before the bounce and in an expanding phase after the bounce). Thus, the quantum corrections to Einstein's equations in the Planck regime are quite subtle, whence the quantum dynamics in the Planck regime has features that could not have been a priori anticipated. 

\bfig
 \ig[width=0.47\textwidth]{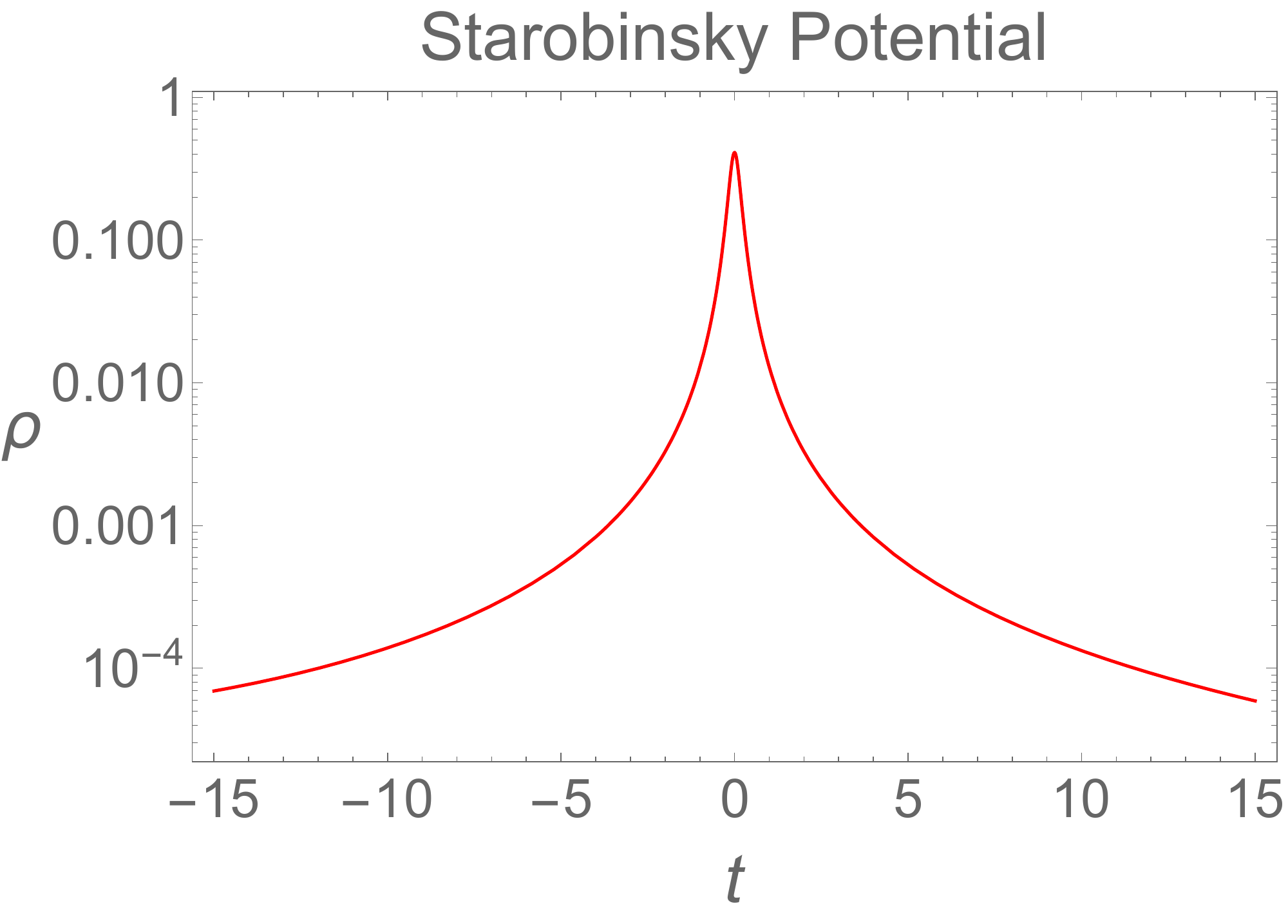}
 \hskip0.5cm
 \ig[width=0.47\textwidth]{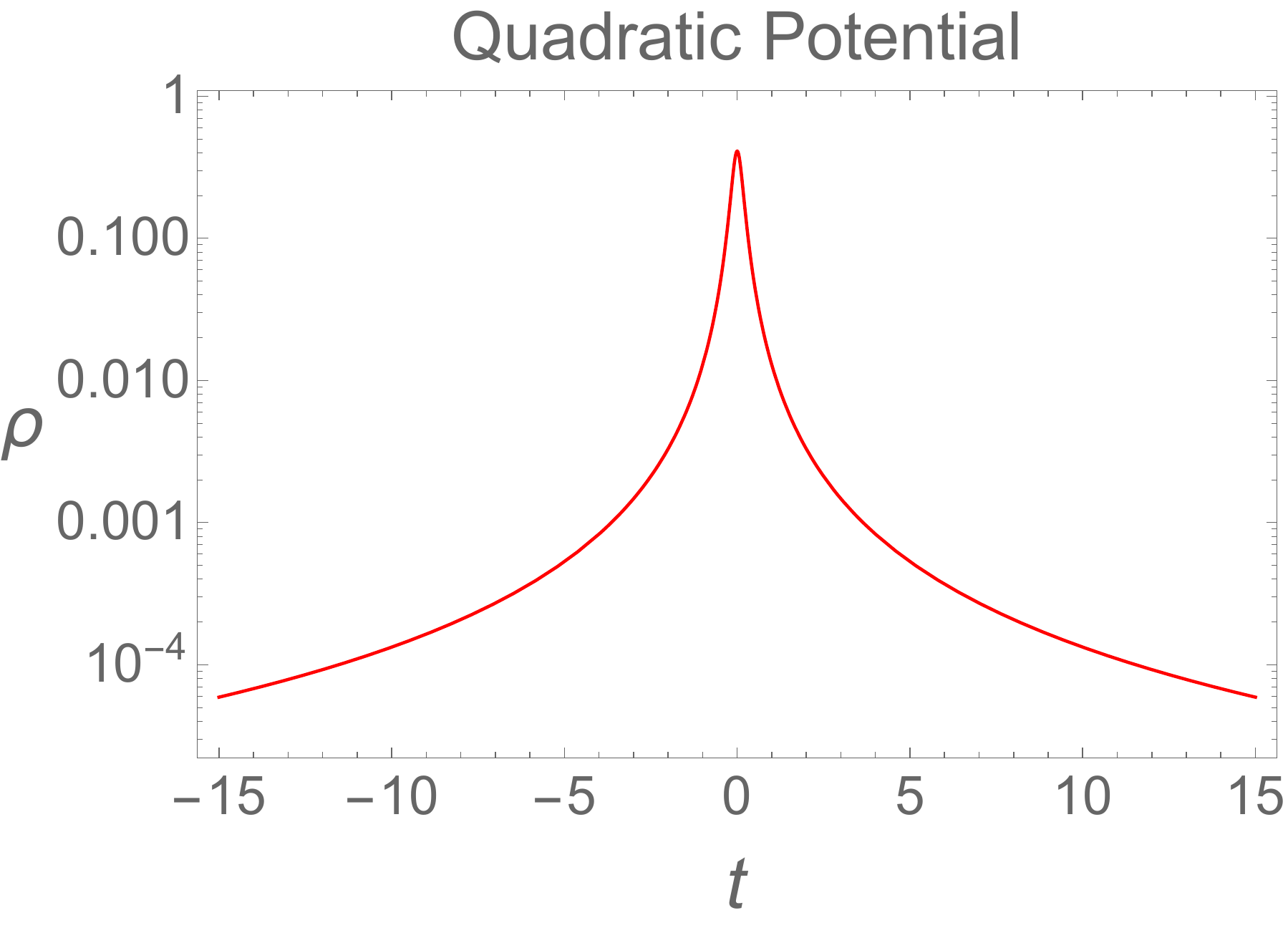}
\caption{Evolution of the matter density in the inflationary scenario, 
in a vicinity of the LQC bounce. x-axis shows proper time in Planck seconds and the y-axis shows matter density in units of Planck density. We use a logarithmic scale on the y-axis because the density falls by 4 orders of magnitude in $\sim\! 11.52$ Planck seconds. (These simulations use initial conditions that are selected by a principle discussed in \cite{ag3}.)\\
{\it Left panel:} The Starobinsky potential, where the mass parameter is fixed using PLANCK data.\\
{\it Right panel:} The quadratic potential, where the mass parameter is fixed using PLANCK data.}
\label{figs:density}
\vskip0.6cm
\efig

One of these features plays an important role in the quantum generalization of the WCH: It enables us to characterize what we will mean by the Planck regime. Once the matter density (or curvature) falls below $10^{-4}\, \rhopl$ the quantum corrections to Einstein's equations become negligible and general relativity becomes an excellent approximation for the finitely many degrees of freedom captured by the mini-superspace. For example, even though the quantum corrected dynamical trajectory is still different from the GR trajectory, the GR trajectory is now within the half-width of the sharply peaked quantum state $\Psi_{o}$. Therefore, by \emph{Planck regime} we will mean the portion of quantum corrected space-time in which the matter density (or curvature) is greater than $10^{-4}$ in Planck units. How long does this phase last, as measured by proper time of effective metric $\underbar{g}_{ab}$ that captures the leading quantum corrections? \fref{figs:density} illustrates that the duration of this phase can be astonishingly short. The plots show the evolution of matter density as a function of proper time defined by the effective metric $\underbar{g}_{ab}$ around the LQC bounce for the Starobinsky and quadratic potentials in the inflationary scenario. In both cases, the density (and curvature) drops by 4 orders of magnitude within less than $\sim 12$ Planck seconds. By contrast, the onset of the slow roll inflation (characterized by the time at which the pivot mode exists the Hubble radius) occurs some $10^{6} - 10^{7}$ Planck seconds later and inflation itself also lasts about $10^{6} - 10^{7}$ Planck seconds. While the precise duration is irrelevant to the statement of our QHIH, the fact that it is so short simplifies the task of actually extracting the small ball $\B$ of states it selects in concrete examples (such as the ones discussed in \cite{ag3}).

We will conclude this discussion by noting a conceptually important point about cosmological perturbations in the Planck regime. Recall from section \ref{s1} that the system under consideration consists of FLRW backgrounds \emph{together with} first order cosmological perturbations thereon. Therefore, we are led to consider states $\Psi_{o}\otimes \psi$ where $\Psi_{o}$ denotes, as before, the state of the quantum FLRW geometry, and $\psi$, the state of a quantum field $\h{h}_{ab}$ representing a cosmological perturbation on $\Psi_{o}$. At first, it seems very difficult to study the dynamics of $\h{h}_{ab}$ on the \emph{quantum} space-time $\Psi_{o}$. However, there is an unforeseen simplification \cite{akl,aan1,aan2}: So long as back-reaction of perturbations $\h{h}_{ab}$ on the quantum FLRW geometry $\Psi_{o}$ can be neglected, the evolution of $\h{h}_{ab}$ on $\Psi_{o}$ is \emph{completely equivalent} to its evolution on a quantum corrected, smooth FLRW metric $\t{g}_{ab}$ which is systematically constructed from $\Psi_{o}$ in a specific way.%
\footnote{That the back-reaction can be neglected is a non-trivial assumption. In practice one makes the assumption, selects states $\Psi_{o}\otimes\psi$ using some principle, calculates $\t{g}_{ab}$ using $\Psi_{o}$, evolves the quantum perturbation $\h{h}_{ab}$ on the background metric $\t{g}_{ab}$ and, in the end verifies that the initial assumption on back reaction is satisfied in this solution. This self consistency check was performed in the initial LQC simulations \cite{aan3}.} 
Thus, all the quantum fluctuations in $\Psi_{o}$ to which the evolution of $\h{h}_{ab}$ is sensitive are distilled in $\t{g}_{ab}$. As one would expect, $\t{g}_{ab}$ is a FLRW metric --but with coefficients that depend on $\hbar$. It captures more information about the quantum state $\Psi_{o}$ than the effective metric $\underbar{g}_{ab}$ defined by the dynamical trajectory on which $\Psi_{o}$ is peaked. However, for sharply peaked states $\Psi_{o}$, $\t{g}_{ab}$ is practically indistinguishable from $\underbar{g}_{ab}$ \cite{ag1}. The FLRW metric $\t{g}_{ab}$ is referred to as the \emph{dressed effective metric}; it is `dressed' by the fluctuations in $\Psi_{o}$ to which the dynamics of perturbations is sensitive.

In what follows we will use the term `quantum FLRW geometry' to refer either to $\Psi_{o}$ \emph{or} the dressed effective metric $\t{g}_{ab}$ extracted from it. The formulation of the quantum generalization of WCH is significantly simplified by the interplay between $\Psi_{o}$ and $\t{g}_{ab}$ \emph{since we can focus only on cosmological perturbations on the background metric} $\t{g}_{ab}$.\\  

\emph{Remark:} There is, of course, a small ambiguity in the characterization of the `Planck regime'. It will trickle down to the ball $\B$ selected by the generalized WCH in concrete models of the early universe. Should we be concerned by this `fuzziness'? If the goal were to single out `the' state that describes the universe in complete detail, one would want a sharp statement without such fuzziness. But recall that in cosmology one is interested in describing only the large scale structure of the universe. Thus, the states we select are \emph{not} meant to capture all the details of space-time geometry, but only the large scale structure resulting from appropriate coarse graining. Given that there are inherent ambiguities in coarse graining, we believe that it is appropriate to have a degree of fuzziness in the statement of the generalized WCH. In particular, if one were to change the precise meaning of the Planck regime somewhat, the states in the new ball $\B$ will still continue to be very special; they will continue to embody the spirit of the original WCH hypothesis of Penrose's. 

\section{Generalized Weyl Curvature hypothesis}
\label{s3}

The system under consideration consists of quantum fields representing first order perturbations, propagating on a quantum FLRW background $\Psi_{o}$. We wish to constrain the states\, $\Psi_{o}\otimes\psi$\, in the Hilbert space of this combined system via a suitable generalization of the WCH. States $\Psi_{o}$ of the background quantum geometry refers to the FLRW mini-superspace on which the Weyl tensor vanishes identically. Therefore the generalized WCH will restrict only states $\psi$ of perturbations. In this section we will first analyze tensor modes and use the final results to arrive at a statement of the QHIH for scalar modes. 

In the present setting, the original WCH would require that the perturbed Weyl tensor of these tensor modes should vanish at the big bang. As explained in section \ref{s1}, we wish to replace the big bang by the Planck regime of the quantum FLRW geometry $\t{g}_{ab}$ and the Weyl tensor by suitable operators. The phase space structure discussed in section \ref{s2.1} leads us to replace the classical Weyl tensor by phase space observables $\Rthree_{\vk}$ and $B_{\vk}$. The vanishing of these observables for all $\vk$ is necessary and sufficient for tensor perturbations to vanish, i.e., to ensure homogeneity and isotropy to first order, beyond the background geometry. But the Poisson brackets (\ref{RBPB}) imply that the corresponding operators satisfy the following `canonical commutation relations' (CCR): 
\be \label{CCR1} [\hRone_{\vk}\,,\,\, \hBone_{\vk^{\prime}}] = i\hbar\, \f{\kappa}{a^{2}}\,\, k^{3}V_{0}\,\, \delta_{\vk,\, -\vk^{\prime}} \, .\ee
Since the right side is a non-zero constant, there is \emph{no} state in the Hilbert space of tensor modes which is annihilated by either of these operators. Thus, because quantum fluctuations cannot be removed even in principle, the notion of quantum homogeneity and isotropy (QHI) is more subtle than that in the classical theory.

The central idea is to restrict the Heisenberg state of tensor modes by demanding that, in the Planck regime\\
(i) the expectation values of both $\hRone_{\vk}$ and $\hBone_{\vk}$ should vanish for all $\vk$;\,\, \\
(ii) the product of uncertainties be minimum; and,\,\, \\
(iii) the uncertainties be equally divided between them.\\
(It is meaningful to impose the last condition because the two curvature operators have the same physical dimensions.) We will find that we will have to sharpen the three conditions appropriately to implement this idea.

We will impose these conditions in two steps. In the first, carried out on section \ref{s3.1}, we implement the ideas in the simpler setting of flat space-time using the framework developed in section \ref{s2.1}. Because there is no Plank regime in flat space-time, this discussion should be thought of only as a mathematical step, but one  that is technically important in order to streamline calculations. (It also highlights the fact that the Fock vacuum is uniquely selected by conditions (i) - (iii).) In the second step, carried out in section \ref{s3.2}, we use the flat space-time results together with the notion of the Planck regime introduced in section \ref{s2.2} to formulate the desired quantum generalization of the WCH. We will find that the dynamical nature of the FLRW background geometry makes it necessary to introduce additional new elements in the passage from the WCH to the final QHIH.

\subsection{Quantum theory: Flat background}
\label{s3.1}

In this subsection we will carry out the first of the two steps mentioned above using the space-time metric $\mathring{g}_{ab}$ of section \ref{s2.1}. This mathematical detour will bring out an interesting fact about linearized gravitational waves in flat space-time: \emph{the quantum vacuum state can be mathematically characterized  using Heisenberg uncertainties between gauge invariant curvature observables}, without direct reference to Poincar\'e invariance or positivity of energy. 

Because we are restricting ourselves to flat space-time, the scale factor is now time independent ($a(\eta) =1$), tensor modes satisfy $\phi^{\prime\prime}_{\vk} + k^{2} \phi_{\vk} =0$, and, the Hamiltonian does not have any explicit time dependence. Therefore we can work just with the Phase space $\ps$  in place of the extended phase space $\psE$. 
The right side of the Poisson brackets between $\Rthree_{\vk}$ and $B_{\vk}$ is also time independent now and hence we have the commutation relations
\be [\h{\Rthree}_{\vk}\,,\,\, \h{B}_{\vk^{\prime}}] = i\,(\hbar\kappa)\,\, (k^{3}V_{0})\,\, \delta_{\vk,\, -\vk^{\prime}} \, .\ee
Unfortunately we cannot directly use these CCR directly in our considerations of uncertainty relation because neither of these curvature operators is self-adjoint: ${\h{\Rthree}}^{\dag}_{\vk} =\h{\Rthree}_{-\vk}$ and $\h{B}^{\dag}_{\vk} = \h{B}_{-\vk}$. Let us therefore make a small detour to introduce self-adjoint linear combinations of the field operators $\h\phi_{\vk}$ and their momenta $\h\pi_{\vk}$:
\ba \label{qp} \h{Q}_{1} &=& \f{a\sqrt{k}}{\sqrt{2\hbar V_{0}}}\, \big(\h\phi_{\vk} + \h\phi_{-\vk}\big) \qquad {\rm{and}} \qquad \h{P}_{1} = \f{1}{a\sqrt{2 k\hbar V_{0}}}\, \big(\h\pi_{\vk} + \h\pi_{-\vk}\big)\nonumber\\
 \h{Q}_{2} &=& \f{a\sqrt{k}}{i\sqrt{2\hbar V_{0}}}\, \big(\h\phi_{\vk} - \h\phi_{-\vk}\big) \qquad {\rm{and}} \qquad \h{P}_{2} = \f{1}{ia\sqrt{2k\hbar V_{0}}}\, \big(\h\pi_{\vk} - \h\pi_{-\vk}\big)\, ,\ea
where we have retained the scale factor $a$ although it equals $1$ in flat space because these expressions will be used also in the next subsection where we consider general FLRW space-times.%
\footnote{That discussion will clarify the reason behind the specific way the scale factor enters in (\ref{qp}).}
In (\ref{qp}) we have chosen numerical factors to make the $\h{Q}_{I}$ and $\h{P}_{I}$ (with $I=1,2$)  dimensionless so that the commutation relations between them are just
\be \label{pqccr}[\h{Q}_{I}, \, \h{P}_{J}] \, = \, i\, \delta_{I,J}\quad {\rm{with}}\,\, 
I, J = 1,2\, . \ee
As noted in Eqs (\ref{Rthree}) and (\ref{B3}), the gauge invariant curvature operators can be expressed in terms of $\h\phi_{\vk}$ and $\h\pi_{\vk}$ as:
\be \label{RthreeB}\h\Rthree_{\vk} = \sqrt{\kappa}\, k^{2}\, \h\phi_{\vk} \quad {\rm and}\quad \h{B}_{\vk} = \f{\sqrt{\kappa}\, k}{a^{2}}\,\, \pi_{\vk}, . \ee
Hence the $\h{\Rthree}_{\vk}\,,\, \h{B}_{\vk},$ can be expressed as complex linear combinations of observables $\h{Q}_{I},\, \h{P}_{I}$: 
\be \h{\Rthree}_{\vk}\, = \Big(\f{(\kappa\hbar)(k^{3}V_{0})}{2a^{2}}\Big)^{\f{1}{2}}\, \Big(\h{Q}_{1} + i \h{Q}_{2}\Big) \quad {\rm and} \quad \h{B}_{\vk} = \Big(\f{(\kappa\hbar)(k^{3}V_{0})}{2a^{2}}\Big)^{\f{1}{2}}\, \Big(\h{P}_{1} + i \h{P}_{2}\Big)\, .\ee
Thus, we can easily go back and forth between the observables $\h{Q}_{I},\, \h{P}_{I}$ and curvature operators $\h{\Rthree}_{\vk}\, ,\h{B}_{\vk}$ of direct interest. After this detour, we will now set the scale factor $a$ once again equal to $1$ in this sub-section.

Since $\h{Q}_{I}$ and $\h{P}_{I}$ are self-adjoint, they are well suited to discuss uncertainty relations. Using commutation relations (\ref{pqccr}) and the standard textbook argument we conclude that they satisfy the uncertainty relations 
\be (\Delta\h{Q}_{I}) (\Delta\h{P}_{J})\, \ge\, \f{1}{2}\, \delta_{I,J}\, .\ee  
We now wish to search for a state $|\psi\rangle$ in the Fock space in which the expectation values of all four observables vanish:
\be \label{cond1}\langle \h{Q}_{I} \rangle = 0\quad {\rm and}\quad  \langle \h{P}_{I} \rangle =0\, ,\ee
the uncertainties are minimized, and shared equally between $\h{Q}_{I}$ and $\h{P}_{I}$. Eq. (\ref{cond1}) implies that expectation values of the two curvature operators vanish and enable us to combine the last two conditions in a convenient form: they are now satisfied if and only if 
\be \label{cond2} \langle \psi|\h{Q_{I}}^{2}+\h{P_{I}}^{2} |\psi\rangle 
\, = \, 1,\quad I=1,2 \ee
or, equivalently, if and only if the curvature operators satisfy
\be \langle \psi| \big(\h\Rthree_{\vk} \h\Rthree_{\vk}^{\dag} + \h{B}_{\vk} \h{B}_{\vk}^{\dag}\big)|\psi\rangle =  (\hbar \kappa)\, (k^{3}V_{o})\, .\ee
The question is: Are there any states $|\psi\rangle$ in the Fock space of gravitons that satisfy these two conditions at a given instant $\eta=\eta_{0}$ of time and, if so, how many?

In the standard Fock space, the field operators are represented by 
\ba 
\h{\phi}_{\vec{k}}(\eta) &=& \Big(\ak\, \f{e^{-ik\eta}}{\sqrt{2k}}\, +\, \admk\, \f{e^{ik\eta}}{\sqrt{2k}} \Big),\nonumber \\ 
 \h{\pi}_{\vec{k}}(\eta) &=& i \Big(- \ak\, \sqrt{\f{k}{2}}\,e^{-ik\eta}\, +\,  \admk\, \sqrt{\f{k}{2}}\, e^{ik\eta}  \Big), 
\label{rep}
\ea
where the creation and annihilation operators are \emph{time independent} and satisfy
  \be [\h{a}_{\vk},\, \h{a}^{\dag}_{\vk^{\prime}}]\, = \,\hbar\, V_{0}\, \delta_{\vk,\vk^{\prime}}\, . \ee
Therefore it is now straightforward to work out the consequences of the two conditions (\ref{cond1}) and (\ref{cond2}). They imply that the state $|\psi\rangle$ must satisfy
\be \langle \psi|[ 1 + \h{N}_{\vk} + \h{N}_{-\vk} ]|\psi\rangle = 1\quad \hbox{\rm for all}\,\,\, \vk \, ,\ee
where $\h{N}_{\vk}$ is the number operator associated with the mode $\vk$. The number operators have a non-negative spectrum and their expectation value vanishes only in the Fock vacuum. Therefore the only state in the Fock space that satisfies our requirement that the homogeneity and isotropy be preserved also by the first order perturbations at a given time $\eta_{0}$ singles out the Fock vacuum uniquely. 

Note that the application of the QHIH at just one instant of time $\eta=\eta_{0}$ has several interesting consequences that could not have been foreseen immediately: (i) the requirement is automatically satisfied at all instants of time $\eta$; (ii) it suffices to select a unique state in the Full Fock space; (iii) although the condition makes no reference to isometries of the metric $\mathring{g}_{ab}$, the state is invariant under the induced action of all isometries; and (iv) is the ground state of the Hamiltonian generating the time translation of $\mathring{g}_{ab}$.\\

\emph{Remark:} We can rewrite Eqs. (\ref{cond1}) and (\ref{cond2}) that encapsulate QHI in terms of the observables $\phi_{\vk}$ and $\pi_{\vk}$ corresponding the scalar field and its conjugate momentum as:
\be \label{cond3} \langle \psi| \h\phi_{\vk}(\eta_{0}) |\psi \rangle =0 \quad {\rm and} \quad  
    \langle \psi| \h\pi_{\vk}(\eta_{0}) |\psi \rangle =0 \ee
and
\be \label{cond4} \langle \psi| \big(k\,\h\phi_{\vk} \h\phi_{\vk}^{\dag} +\f{1}{k} \h{\pi}_{\vk} \h{\pi}_{\vk}^{\dag}\big) (\eta_{0})|\psi\rangle =  \hbar\,\,V_{o}\, .\ee
Therefore, if we were interested in scalar rather than tensor perturbations in flat space, we can use Eqs. (\ref{cond3}) and (\ref{cond4}) to impose the QHI requirement  (The multiplicative factors involving $k$ arise directly from (\ref{cond2}), as they must must because of the difference in the physical dimensions of $\h\phi_{\vk}$ and $\h\pi_{\vk}$.) Again, these conditions select a unique state $\psi$ which is just the vacuum state on the Fock space of scalar perturbations. Thus, a natural extension of the WCH from tensor perturbations leads to QHI conditions also for scalar perturbations.

\subsection{Quantum theory: The Planck regime}
\label{s3.2}

Let us return to the physical system of interest: quantum perturbations propagating on a quantum FLRW background geometry defined by the dressed effective metric $\t{g}_{ab}$. Again, we will first analyze tensor modes and use those results to arrive at a QHIH for scalar modes. In the first part \ref{s3.2.1} of this subsection we will fix a time $\eta_{0}$ in the Planck regime of $\t{g}_{ab}$ discussed in section \ref{s2.2} and ask the Heisenberg state to satisfy the three conditions that encapsulate the QHI requirement. While the mathematical steps used above in flat space will continue to play a key role, two new elements will arise because now the space-time metric is $\t{g}_{ab}$, which is time dependent. First, unlike in section \ref{s3.1}, we do not have a canonical Fock representation of the observable algebra; there is  freedom to perform the Bogoliubov transformations. Therefore, the setting in section \ref{s3.2.1} will be more general than that in section \ref{s3.1}. Second, since the right side of (\ref{CCR1}) is now time dependent, the right hand side of the necessary generalization of (\ref{cond4}) will also carry time dependence. Therefore, the state selected by the new conditions at a given time $\eta_{0}$ in the Planck regime will not satisfy that condition at another time in this regime. Consequently, we will find in section \ref{s3.2.2} that the QHIH now selects a preferred ball in the space of states, rather a unique state. Any state in the ball will be `as homogeneous and isotropic as allowed by the Heisenberg uncertainties and quantum dynamics in the Planck regime'.

\subsubsection{Imposing the QHIH at a given time $\eta_{0}$}
\label{s3.2.1}

Since we no longer have a canonical Fock representation, we need to specify the initial class of candidate states on which to impose the QHI requirements. Recall that in FLRW space-times one constructs a representation of the CCR by choosing a `complete positive frequency basis' $q_{k}(\eta) e^{i\vk\cdot\vx}$ in the space of solutions to the Klein Gordon equation, normalized via 
\be \label{normalization} \big(q_{k}q_{k}^{\prime\star} \, -\, q^{\star}_{k} {q^{\prime}}_{k} \big) (\eta) = \f{i}{a^{2}(\eta)}\, , \ee
and represent $\h{\phi}_{\vk}, \, \h{\pi}_{\vk}$ in terms of the creation and annihilation operators, associated with these basis functions:
\be \label{phipi} \h\phi_{\vk}(\eta)\, = \, q_{k}(\eta)\,\,\h{a}_{\vk}\, +\, q_{k}^{\star}(\eta)\,\, \h{a}^{\dag}_{-\vk} \quad {\rm and} \quad
\h\pi_{\vk}(\eta) = {a^{2}}(\eta)\,\, \big(q^{\prime}_{k}(\eta)\,\,\h{a}_{\vk} + (q^{\prime}_{k})^{\star}(\eta)\,\,\h{a}_{-\vk}^{\dag} \big)\, \, .\ee
We will restrict ourselves to states that belong to one of these infinitely many Fock representations. These states are sometime referred to as \emph{quasi-free states} \cite{quasi-free}.

Now, given any real function $\mu(k)$, and a set $\{q_{k}(\eta)\}$ of permissible basis functions, $\{e^{i\mu(k)}q_{k}(\eta)\}$ is also a set of permissible basis functions. Furthermore, it defines the same Fock representation of the CCR because the state $|0\rangle$ is annihilated by $\h{a}_{\vk}$ if and only if it is annihilated by $e^{i\mu(k)}\, \h{a}_{\vk}$. The invariant structure that characterizes any one Fock representation  is a complex structure $J$ on the classical phase space that is compatible with the natural symplectic structure $\Omega$ thereon \cite{aa-am}.%
\footnote{The complex structure $J$ enables one to decompose any real solution $\phi$ to the linear field equations into a positive and a negative frequency part via $\phi_{\pm} = (1/2)(\phi \mp i J\phi)$. The
compatibility condition is that the second rank tensor $g$ defined by $g(.,.) = \Omega(.,J.)$ is a positive definite metric on the phase space, i.e., that the triplet $(\omega, J, g)$ endows the phase space with the structure of a K\"ahler space. This characterization of a Fock representation holds for any boson field on general, globally hyperbolic curved space-time, not just FLRW cosmologies.} 
In the FLRW space-times now under consideration, the bases $\{ q_{k}\}$ and $\{ e^{i\mu(k)}q_{k}\}$ define the same complex structure. We will make use of this fact in what follows.
 
To impose the QHI requirement, let us begin with the CCR (\ref{CCR1}). Again, we cannot use these CCR directly to analyze consequences of the uncertainty principle because the operators $\h\Rthree_{\vk},\, \h{B}_{\vk}$ are not self-adjoint. Therefore, as in section \ref{s3.1} we are led to define self-adjoint operators $\h{Q}_{I},\, \h{P}_{I}$ of Eq. (\ref{qp}). Again, we seek states $|\psi\rangle$ in which the expectation values of $\h{Q}_{I}$ and $\h{P}_{I}$ vanish (Eq (\ref{cond1})) and the uncertainties satisfy 
\be \label{cond5} \big(\Delta \h{Q}_{I}\big)\big(\Delta \h{P}_{I}\big) = \f{1}{2} \quad {\rm and} \quad \Delta \h{Q}_{I} = \Delta \h{P}_{I}\, . \ee
at a given instant of time $\eta=\eta_{0}$. The differences from section \ref{s3.1} are that these conditions are imposed on the class of all quasi-free states, rather than those belonging a specific Fock representation, and the slice $\eta=\eta_{0}$ is now assumed to be in the Planck regime. We note that conditions (\ref{cond1}) and (\ref{cond5}) are equivalent to assuming that the expectation values of the curvature operators are zero: 
\be \langle \h\Rthree_{\vk}(\eta_{o}) \rangle =0\quad {\rm and}\quad \langle \h{B}_{\vk} (\eta_{o}) \rangle =0\, ,\ee 
and their dispersions are minimal and equally distributed: 
\be \label{cond6}
|\Delta \h\Rthree_{\vk}|(\eta_{0})\,\, |\Delta \h{B}_{\vk}|(\eta_{0})\, =\, \f{(\kappa\hbar) (k^{3}V_{0})}{a^{2}(\eta_{0})}\,  \quad {\rm and} \quad |\Delta \h\Rthree_{\vk}|(\eta_{0})\, = \, |\Delta \h{B}_{\vk}|(\eta_{0})\, \ee
where the dispersions are given by
\be |\Delta \h\Rthree_{\vk}|^{2} = \langle \psi|\h\Rthree_{\vk}\h\Rthree^{\dag}_{\vk}|\psi\rangle \quad {\rm and} \quad  |\Delta \h{B}_{\vk}|^{2} = \langle \psi|\h{B}_{\vk}\h{B}^{\dag}_{\vk}|\psi\rangle \ee
(since  $\langle \h\Rthree_{k}(\eta_{o}) \rangle =0$ and $\langle \h{B}_{k} (\eta_{o}) \rangle =0$). These are precisely the conditions of our QHIH. \emph{The specific placement of the scale factor $a$ in the definition (\ref{qp}) of $\h{Q}_{I}$ and $\h{P}_{I}$ is necessary and sufficient to ensure that the desired conditions (\ref{cond6}) on curvature operators results from the conditions (\ref{cond1}) and (\ref{cond5}) on the self-adjoint operators $\h{Q}_{I}, \h{P}_{I}$.}

Thus, to formulate the QHIH, we can work with the self-adjoint operators $\h{Q}_{I}$ and $\h{P}_{I}$. The standard treatments of the uncertainty principle imply that condition (\ref{cond5}) restricts $|\psi\rangle$ to be a coherent state (in the Hilbert space $\mathcal{H}_{1}\otimes\mathcal{H}_{2}$ corresponding to $\h{Q}_{I}, \h{P}_{I}$\,).  Condition (\ref{cond1}) then restricts $|\psi\rangle$ to be the \emph{unique} coherent state peaked at $\langle \h{Q}_{I} \rangle =0$ and $\langle \h{P}_{I} \rangle$=0. That is, the two conditions together imply that 
\be \big(\h{Q}_{I} + i \h{P}_{I}\big) |\psi\rangle =0, \quad I=1,2. \ee
It is straightforward to translate this requirement in terms of $\h\phi_{\vk}$ and $\h\pi_{\vk}$\,: the state $|\psi\rangle$ must satisfy 
\be \label{key} \Big( \sqrt{k} a\, \h\phi_{\pm\vk}\, +\, i \f{1}{a\sqrt{k}}\, \h\pi_{\pm \vk}\Big) |\psi\rangle =0 \, . \ee 
Recall that, because of the freedom to perform Bogoliubov transformations, we have infinitely many Fock representations of the CCR, each characterized by a (permissible) complex structure $J$. Thus, to begin with $|\psi\rangle$ can be a state in any of these Fock spaces $\mathcal{F}_{J}$. To explore the restriction imposed by (\ref{key}), let us rewrite this condition using creation and annihilation operators that feature in the representation (\ref{phipi}) of $\h\phi_{\vk}$ and $\h\pi_{\vk}$ on the Fock space $\mathcal{F}_{J}$, selected by some basis functions $q_{k}(\eta)$. It turns out that the  expression in terms of creation and annihilation operators is particularly transparent for one specific complex structure $J$. As one could have anticipated from our discussion of section \ref{s3.1}, this $J$ is associated the basis functions $q_{k}(\eta)$ with initial data   
\be \label{qk} q_{k}(\eta_{0}) = \f{e^{-ik\eta_{0}}}{\sqrt{2k}\, a(\eta_{0})}\quad {\rm and}\quad  q^{\prime}_{k}(\eta_{0}) = -i\, \f{\sqrt{k}}{\sqrt{2}}\, \f{e^{-ik\eta_{0}}}{a(\eta_{0})}\, . \ee
at time $\eta_{0}$ (or, more generally, by the basis functions $e^{i\mu(k)}\,q_{k}(\eta)$ for some real functions $\mu(k)$). Simple algebra shows that in this representation, conditions (\ref{key}) on $|\psi\rangle$ are equivalent to requiring simply 
\be \h{a}_{\pm\vk}\, |\psi\rangle =0\, .\ee
Thus, through an inspired guess, we have obtained the Heisenberg state 
$|\psi\rangle$ that satisfies QHI at time $\eta=\eta_{o}$: It is 
the vacuum state $|0_{J}\rangle$ in the Fock representation of the canonical algebra that is selected by the complex structure $J$, picked out by the basis functions (\ref{qk}). 

What would happen if were to we begin with another complex structure $\t{J}$ selected by an inequivalent set of basis functions $\t{q}_{k}(\eta)$? Suppose that the tilde representation is unitarily equivalent to the one selected by $J$, i.e., there exists a unitary operator $U: \mathcal{F}_{J} \to \mathcal{F}_{\t{J}}$ such that the `concrete' operators representing $\h{\phi}_{k}$ and $\h{\pi}_{k}$ via (\ref{phipi}) on the Fock space $\mathcal{F}_{J}$ are mapped to the corresponding `concrete' operators representing them in the tilde representation.%
\footnote{This is the case if $(J-\t{J})$ is Hilbert Schmidt on the 1-particle subspace of $\mathcal{F}_{J}$ \cite{aa-am2}, or, equivalently if the Bogoliubov transformation relating the two bases, $\t{q}_{k}(\eta) = \alpha q_{k}(\eta) + \beta q_{k}^{\star}(\eta)$, is such that $\sum_{k}\,|\beta|^{2}\, < \infty$.}
Then the solution to (\ref{key}) is the unique state $|\psi_{\t{J}}\rangle$ in $\mathcal{F}_{\t{J}}$ which is the image of the vacuum state in the untilde representation: $|\psi_{\t{J}}\rangle = U|0_{J}\rangle$. As is well-known this is a squeezed vacuum in $\mathcal{F}_{\t{J}}$. 
If the tilde representation is not unitarily equivalent to the one determined by $J$, then there is no state in $\mathcal{F}_{\t{J}}$ that will satisfy Eq (\ref{key}), i.e., meet the QHI requirement (\ref{cond6}).

To summarize, up to unitary equivalence, the imposition of QHIH conditions  (\ref{cond1}) and (\ref{cond5}) at a time $\eta_{0}$ in the Planck regime selects a unique quasi-free state on the canonical algebra of operators $\h\phi_{\vk}, \, \h\pi_{\vk}$, or, equivalently, $\h\Rthree_{\vk},\, \h{B}_{\vk}$. This is the vacuum state $|0\rangle$ in the Fock representation selected by the basis functions (\ref{qk}).
\\

\emph{Remarks}:

1. Although we did not explicitly impose any symmetry requirements on $|\psi\rangle$, our final result implies that it is automatically invariant under the induced action of spatial symmetries of the FLRW metric $\t{g}_{ab}$. This follows from considerations of the 2-point function \cite{aan2}. In the vacuum state of the Fock representation selected by \emph{any} basis functions $q_{\vk}(\eta)$, it has the form
\be \langle 0|\h{\phi}(\vx,\eta)\, \h{\phi}(\vx^{\prime}, \eta^{\prime})|0\rangle = \f{\hbar}{V_{0}}\, \sum_{\vk}\,\Big( q_{k}(\eta)\,q_{k}^{\star}(\eta^{\prime})\Big) e^{i\vk\cdot\vx}\,\, . \ee
Since the right side is manifestly invariant under the action of spatial isometries of $\t{g}_{ab}$, and since the 2-point function suffices to characterize the vacuum, this implies the invariance of all these vacua $|0\rangle$. We will denote the collection of all these vacua by $\mathcal{C}$ and refer to them as \emph{weakly homogeneous and isotropic states}. Because our $|\psi\rangle$ is one such vacuum state, it follows immediately that it belongs to the collection $\mathcal{C}$.

2. The above remark makes it clear that there are \emph{infinitely many} (unitarily inequivalent) states that are weakly homogeneous and isotropic, one for each choice of permissible complex structure $J$. The QHIH conditions (\ref{cond1}) and (\ref{cond5}) (or, equivalently, (\ref{cond6})) are much stronger: they select a \emph{unique} state $|\psi\rangle$ (up to unitary equivalence). In particular, the vacuum state in the Fock representation determined by basis functions which determine a \emph{different} complex structure $J$ than the one selected by the $q_{k}(\eta)$ of Eq. (\ref{qk}) do \emph{not} satisfy the QHIH at time $\eta_{0}$.

3. Results we just obtained in section \ref{s3.2} imply that even in flat space-time of section \ref{s3.1} we need not have started with the standard Fock representation. We could have begun with general quasi-free states and \emph{arrived at} the standard Fock vacuum through the QHI requirement.

4. However, there is a key difference between the situation in flat space-time and the general FLRW space-times that we now consider. Although we began our analysis with a fixed time $\eta_{o}$ also in flat space-time, as we saw in section \ref{s3.1}, the resulting state is not only unique but the \emph{same} for all choices of $\eta_{0}$. As explained below, this is no longer the case in a general FLRW space-time. 

\subsubsection{Imposing the QHIH in the full Planck regime} 
\label{s3.2.2}

Recall from section \ref{s2.1} that by \emph{Planck regime} we mean the portion of the quantum space-time in which the matter density (or curvature) is greater than $10^{-4}$ in Planck units. Let us foliate this portion of space-time by cosmic slices and suppose that the conformal time $\eta$ corresponding to these slices lies in a closed interval $I$. Let us choose a time $\eta_{0} \in I$ and denote the Heisenberg state satisfying the QHIH at this time by $|0_{\eta_{0}}\rangle$. In this state the expectation values of $\h\Rthree_{k}$ and $\h{B}_{k}$ vanish and 
\be \label{cond7} \langle 0_{\eta_{0}}| \Big(\h\Rthree_{\vk} \h\Rthree_{\vk}^{\dag} + \h{B}_{\vk} \h{B}_{\vk}^{\dag}\Big) (\eta_{0})|\,0_{\eta_{0}}\rangle\, =\,  \f{(\kappa\hbar)\,(k^{3}V_{0})}{a^{2}(\eta_{0})}\, . \ee
The right hand side represents the minimum that the expectation value can achieve in any quasi-free state. Hence, at a different time $\eta \in I$ we will generically find  
\be \label{inequality} \langle 0_{\eta_{0}}| \Big(\,\h\Rthree_{\vk} \h\Rthree_{\vk}^{\dag} + \h{B}_{\vk} \h{B}_{\vk}^{\dag}\Big) (\eta)\,|0_{\eta_{0}}\rangle \,\, >  \,\, \f{(\kappa\hbar)\,(k^{3}V_{o} )}{a^{2}(\eta)}\, . \ee
Since the expectation values of $\h\Rthree_{k}$ and $\h{B}_{k}$ will continue to vanish also at time $\eta$, the left side of (\ref{inequality}) continues to be a faithful measure of uncertainties, whence (\ref{inequality}) captures the fact that in general uncertainties in curvature operators will grow. Therefore there is \emph{no state} which will minimize the uncertainties for all times in the full Planck regime. Is there a more appropriate measure of minimum uncertainties associated with the full Planck regime that we can use?  To analyze this issue, first note that we can monitor how uncertainties change in time in the Planck regime. Given any two times $\eta_{1}, \eta_{2}$ in the Interval $I$, consider the quantity
\be \sigma_{k}^{2} ({\eta_{1}},\,{\eta_{2}}) := \langle 0_{\eta_{1}} |\, \Big(\,\h\Rthree_{\vk} \h\Rthree_{\vk}^{\dag} + \h{B}_{\vk} \h{B}_{\vk}^{\dag}\Big) (\eta_{2})\,|0_{\eta_{1}} \rangle \, .\ee
Since the interval $I$ is compact and $a(\eta)$ is bounded away from zero, $\sigma_{k}^{2} ({\eta_{1}},\,{\eta_{2}})$ is bounded above as we vary $\eta_{1}, \eta_{2}$ in the interval $I$. Let us set 
\be \label{doublesup} s_{k}^{2}\, :=\,  \sup_{\eta_{1},\eta_{2}\in I}  \sigma_{k}^{2} \, ({\eta_{1}}\, ,{\eta_{2}})\, . \ee
This quantity provides us a definite measure of minimum dispersions we must allow if we wish to characterize states that are preferred by uncertainty considerations in the full Planck regime, i.e. the full interval ${I}$. 

These considerations suggest the following strategy. Let us start with the collection $\mathcal{C}$ of all \emph{weakly} homogeneous and isotropic states. Up to unitary equivalence, each of these states is a vacuum $|0_{J}\rangle$ for some choice of permissible complex structure $J$ and hence satisfies 
\be \label{cond8}\langle 0_{J}| \h\Rthree_{k}(\eta)| 0_{J}\rangle = 0 \quad{\rm and} \quad \langle 0_{J}| \h{B}_{k}(\eta)| 0_{J}\rangle = 0, \quad \forall \eta\, . \ee
However, dispersions in the curvature operators can be \emph{arbitrarily large} at any given time $\eta\in I$ in the Planck regime. Thus a generic \emph{weakly} homogeneous and isotropic state will not come even close to satisfying the desired \emph{quantum} homogeneity and isotropy.
However, since there is a subset of $\mathcal{C}$ consisting of states $|0_{\eta}\rangle$ which \emph{minimize} the dispersion at the given time $\eta\in I$, at first it seems natural to restrict oneself to just this subset. However, given any one $|0_{\eta}\rangle$, the dispersions are bounded only by $s^{2}_{k}$ at other times. Therefore, \emph{as far as the full Planck regime is concerned,} this 1-parameter family is on the same footing as any other weakly homogeneous and isotropic state that satisfies  
\be \label{ball1} \langle 0_{J}|\, \Big(\h\Rthree_{\vk} \h\Rthree_{\vk}^{\dag} + \h{B}_{\vk} \h{B}_{\vk}^{\dag}\Big) (\eta)\,|0_{J}\rangle\, \le s_{k}^{2}, \quad \forall \eta\in I \, .\ee
Therefore we are led to consider the ball $\B$ in the space $\mathcal{C}$ of weakly homogeneous and isotropic states $|0_{J}\rangle$ satisfying (\ref{ball1}). The entire set $\mathcal{C}$ is much too large for QHI, while the one parameter family $|0_{\eta}\rangle$ (with $\eta\in I$) is too small. The ball $\mathcal{B}$ lies between the two and captures all weakly homogeneous and isotropic states that `come as close to satisfying QHI requirement as the Heisenberg uncertainty principle and quantum dynamics allow in the full Planck regime'. \emph{For tensor perturbations then, our statement of the QHIH is simply that the Heisenberg state should belong to this ball $\B$.}

Let us summarize our discussion of tensor modes. In the full classical theory, Penrose's WCH restricts the initial state by requiring that the Weyl tensor should vanish at the big bang. For the current paradigms of the early universe, we need to extend the underlying idea to select initial conditions for \emph{quantum fields} representing  tensor perturbations, replacing the big bang by the Planck regime. Therefore, we propose that the WCH be replaced by the QHIH:
\begin{quote}
In the Heisenberg picture, only those quasi-free Heisenberg states of tensor perturbations should be allowed which are\\ (i) weakly homogeneous and isotropic, i.e., left invariant by the induced action of the FLRW spatial isometries; and, \\ (ii) in which the dispersions in these curvature operators are as small and as equally distributed as possible in the Planck regime. 

Note that (i) implies, in particular, that condition (\ref{cond8}) is satisfied. The precise restriction on quasi-free states imposed by (iii) is given in Eq.  (\ref{ball1}).  
\end{quote}

Finally, let us consider scalar perturbations. As remarked at the end of section \ref{s3.1}, since the operators $\h\Rthree_{\vk},\, \h{B}_{\vk}$ can be expressed in terms of the `metric perturbations' $\h\phi_{\vk},\, \h\pi_{\vk}$, we can readily carry over these considerations to scalar perturbations. As explained in \cite{aan2,aan3}, the gauge invariant 
Mukhanov-Sasaki scalar perturbations are well defined also during pre-inflationary dynamics.%
\footnote{Curvature perturbations are also gauge invariant and particularly convenient during inflation because their Fourier modes freeze after the mode exists the Hubble horizon. However, they are related to the Mukhanov-Sasaki variables by a factor involving the inverse of the background inflaton field and hence they diverge as the inflaton turns around in the pre-inflationary phase.}
Therefore let us work with these variables. Since they have the physical dimensions of a scalar field, for simplicity of notation, let us denote them just by $\h\phi_{\vk}, \h\pi_{\vk}$. Using Eq. (\ref{cond4}), we define the ball $\mathcal{B}$ as follows: It consists of the those weakly homogeneous and isotropic states $|0_{J}\rangle$ in the Fock representations of the CCR which satisfy 
\be \label{ball2} 
\langle 0_{J}|\, \Big(k\h\phi_{\vk} \h\phi_{\vk}^{\dag} + \f{1}{k}\h{\pi}_{\vk} \h{\pi}_{\vk}^{\dag}\Big) (\eta)\,|0_{J}\rangle\, \le {s^{\prime}}_{k}^{2} \quad \forall \eta\in I  \ee
where ${s^{\prime}}^{2}_{k}$ refers to the supremum as in (\ref{doublesup}), but now obtained using the Mukhanov-Sasaki fields and their conjugate momenta. Then, 
\begin{quote}
The QHIH for scalar modes is the same as that for tensor modes except that (\ref{ball1}) is now replaced by (\ref{ball2}). 
\end{quote}

\emph{Remarks:}\\
1. The set $\{|0_{\eta}\rangle\}$ is distinguished by the fact that in the state $|0_{\eta}\rangle$ the dispersions in the curvature operators are minimum possible at time $\eta$. Furthermore the Fock representations these vacua $|0_{\eta}\rangle$ define are unitarily equivalent (i.e. the Bogoliubov coefficients that relate these mode functions satisfy $\sum_{k} |\beta_{k}|^{2} < \infty$). However, the ultraviolet behavior of these states is not sufficiently regular for the expectation value of the stress-energy tensor operator to be well-defined \cite{ia-aa}. The ball $\B$ on the other hand does contain states which satisfy ultraviolet regularity to 4th (or higher) adiabatic order, ensuring that the expectation value of the renormalized stress-energy tensor is well-defined at all times. We were led to enlarge the set of states of interest from $\{|0_{\eta}\rangle\}$ to $\B$ by the fact that, for considerations of dispersions \emph{in the full Planck regime}, states in the full ball $\B$ are on the same footing as those in the set $\{|0_{\eta}\rangle\}$.  Regularity considerations bring out another facet of the necessity of enlargement. Indeed, if we were to add physical considerations beyond those motivated by the WCH, a natural criterion would be to demand adiabatic regularity to at least order 4 \cite{aan2,aan3}. If this is done, the ball $\B$ would shrink. But the smaller (ultraviolet regular) ball still allows interesting Heisenberg states. For example, in the context of inflation with a quadratic or Starobinksy potential, the ball would contain the Bunch-Davies vacuum selected a few e-folds before the pivot mode exits the Hubble horizon, as well as the so-called ANA-vacuum \cite{ana}, selected by stress-energy considerations near the bounce \cite{ag3}.   

2. Consider the set of all weakly homogeneous and isotropic states $|0_{J}\rangle \in \mathcal{C}$. Using any one of these, $|0_{J_{0}}\rangle$ as the origin, we can coordinatize  any other $|0_{J}\rangle$ using the Bogoliubov coefficients relating them, i.e., by two functions $r_{k}, \theta_{k}$ which span a 2-dimensional plane for each $k$:\, $r_{k} \ge 0$ and $\theta_{k} \in (0,2\pi)$. Thus $\mathcal{C}$ can be endowed the structure of an infinite dimensional vector space. The ball $\B$ is a small subset of this full space in that for each $k$ we now have only a disc $r_{k} \le s_{k}$. But one may wish to impose additional regularity or physical conditions to further narrow down the choice of the Heisenberg states by appealing to some features of quantum perturbations beyond Planck scale dynamics. An example of such a requirement is discussed in \cite{ag3}, where one narrows down the choice to a \emph{unique} Heisenberg state through a requirement that bridges the Planck scale dynamics to physics at the end of inflation.

\section{Summary and Discussion}
\label{s4}

The issue of whether the observed large scale properties of the universe can result from generic initial conditions or whether a past hypothesis is essential for these properties to emerge has been debated for a long time. In this paper we have adopted the second view, following the line of thinking that underlies Penrose's WCH. But we formulated our discussion in the context of the current paradigm of the very early universe which is both a restriction and a generalization of the setting used in the original formulation of the WCH: Rather than using full non-linear classical general relativity, one restricts one's attention only to  FLRW geometries together with linear perturbations, which, however, are now described using \emph{quantum field theory}. The issue of initial conditions persists also in this setting because there is still a great deal of freedom in the choice of the background FLRW geometry as well as the Heisenberg state of quantum fields representing perturbations. However, since Weyl curvature vanishes identically in FLRW space-times, Penrose-type considerations can lead to restrictions on initial conditions only for perturbations. Our discussion was focused on this problem.

To arrive at these restrictions, we had to extend Penrose's original considerations in three directions. First, since initial conditions for quantum fields cannot be specified on a singularity, we are forced to replace the big bang by the Planck regime of a singularity-free quantum gravity theory. Second, to formulate our generalization of the WCH, we had to pay due attention to the uncertainty relations. To obtain these relations, in turn, we had to first express curvature tensors in terms of phase space observables and compute their Poisson brackets, adding a new dimension to the purely space-time considerations of the original WCH. Finally, while Weyl curvature considerations are well-suited for tensor modes, they are not adequate for scalar modes. Therefore a parallel strategy has to be developed for scalar modes.

To carry out the first of these tasks, we were guided by LQC, where the big bang is replaced by a quantum bounce due to quantum geometry effects and quantum fields representing cosmological perturbations propagate on the resulting quantum geometry. As one would expect, at the bounce the curvature and matter density are of Planck scale and general relativity provides a good description of the background FLRW geometry once they fall by a factor of $10^{-4}$. This provides a good control over what one means by the Planck regime. In the actual formulation of our generalized WCH, only the qualitative features of these results are relevant. Therefore, the formulation is applicable well beyond LQC. In carrying out the second task, one would first expect to write uncertainty relations using the electric and magnetic parts of the Weyl tensor, $E_{ab}$ and  $B_{ab}$. However, it turned out that while the magnetic part can be represented by phase space functions, the electric part cannot, because it involves \emph{second} time derivatives of the metric in (full general relativity as well as) the linearized theory. But one can use instead the intrinsic Ricci tensor $\Rthree_{ab}$ on spatial slices (or, equivalently, the extrinsic curvature of these slices). The pair $\Rthree_{ab},\,  B_{ab}$ is canonically conjugate in the sense that the Poisson bracket between them is a c-number (see Eq. (\ref{RBPB})). The CCR (\ref{CCR1}) resulting from these Poisson brackets lead to Heisenberg uncertainty relations for the two sets of observables. Our statement of the generalized WCH asks that the Heisenberg state be so restricted that the expectation values of all these observables be zero and dispersions between them be as small in the full Planck regime as is allowed by the Heisenberg uncertainties and the Planck scale dynamics; see Eqs. (\ref{cond8}) and (\ref{ball1}). Finally, the third task --incorporation of scalar modes-- can be carried out easily by first expressing these equations in terms of the metric perturbations and their conjugate momenta and carrying over that requirement to scalar modes. 

At a technical level, the analysis brought out the powerful role played by considerations involving uncertainty principles. Along the way we saw that these considerations are sufficient to select the vacuum state uniquely in the standard Fock representation in flat space-time, and also select a unique state (up to unitary equivalence) if imposed at any one time in a quantum FLRW geometry. However, because the quantum FLRW geometry is time dependent, there is no state that satisfies (\ref{cond8}) and (\ref{ball1}) in the entire Planck regime. Rather, what we obtain is a small ball $\B$ of Heisenberg states that satisfies the quantum generalization of the WCH in the full Planck regime. These states are as homogeneous and isotropic in the full Planck regime as allowed by the uncertainty principle and quantum dynamics. Therefore, they satisfy the QHIH --quantum homogeneity and isotropy hypothesis-- that generalizes the WCH. The ball $\B$ contains the Heisenberg states that have been used in the literature. Examples are \cite{ag3}: (i) the so-called ANA vacua \cite{ana} in which the expectation value of the adiabatically regularized stress-energy tensor vanishes (mode by mode) at a given instant of time in the Planck regime, and, (ii) in the inflationary paradigm with the quadratic or Starobinsky potentials, the Bunch-Davies vacua. (Other potentials have not been analyzed in detail but we expect the result to be much more general.) However, recall that the Bunch-Davies vacuum is normally selected by appealing to the approximate de Sitter geometry near the onset of inflation. From a conceptual viewpoint, this corresponds to imposing `initial conditions' in the middle of evolution when the curvature is about $10^{-11}$ times the Planck curvature. The ball $\B$, on the other hand, is selected by fundamental uncertainty relations in the Planck regime. One can impose further conditions to reduce the size of the ball $\B$. Examples are: stronger ultraviolet regularity in the Planck regime, and/or a condition that bridges the Planck regime with the late time universe. An example of the second kind that leads one to a unique state in the ball is discussed in \cite{ag3}.   

The proposal requires one to specify the Planck regime. Since this specification can vary because there is no sharp boundary, the precise statement of the QHIH has a certain degree of fuzziness. However, our goal was to specify the initial conditions that would capture only the coarse grained, large scale structure of our universe, rather than its detailed microscopic features. Since the notions of `coarse-graining' and `large scale structures' are themselves fuzzy, we believe that it is reasonable to have the corresponding degree of fuzziness in permissible initial conditions. In specific inflationary models referred to above, numerical simulations have shown that final results for observable modes in the CMB are quite insensitive to the precise specification of the  Planck regime \cite{ag3}. This is just as one would expect because these modes only probe the large scale structure. 

How does the QHIH compare to other ideas such as the Hartle-Hawking (HH) no boundary proposal \cite{hh} aimed at constraining the quantum state of the universe? The underlying idea in the HH proposal can be stated in the context of full quantum gravity. However, to make it precise, one would have to spell out how to perform path integrals. For this, typically one restricts oneself to the mini-superspace approximation for the background. But even in this case, the path integrals are yet to be performed satisfactorily beyond some truncation,  e.g. the steepest descent approximation. Inclusion of perturbations in path integrals has also remained formal. Therefore, it is not yet clear what the precise class of Heisenberg states selected in this approach is. What about QHIH? Like the WCH from which it draws inspiration, the idea underlying QHIH is restricted to the cosmological setting, which, however, can be much more general than the current paradigm of the early universe used in our analysis. But it is not obvious how to extend it to full quantum gravity. Furthermore, the detailed implementation of the idea has been carried out only in the context of quantum perturbations on quantum FLRW geometries. This implementation does not require further truncations; all calculations can be carried out systematically by `lifting' techniques from quantum field theory in classical FLRW geometries to quantum \cite{akl,aan2,aan3,ag3}. As a result, we have full control on the class of states chosen by this principle: states that lie in the ball $\B$. 

In any one of these states the quantum universe is maximally isotropic and homogeneous in the Planck regime. This is the past hypothesis embodied in the QHIH. The idea, as in the WCH, is that the arrow of time emerges from this very special initial condition.

\vskip0.04cm
\acknowledgements{We would like to thank Ivan Agullo and William Nelson for discussions. This work was supported in part by the NSF grant PHY-1505411, the Eberly research funds of Penn State and Perimeter Institute for Theoretical Physics. Research at Perimeter Institute is supported by Government of Canada through the Department of Innovation, Science and Economic Development and by the Province of Ontario through the Ministry of Research, Innovation and Science. This research used the Extreme Science and Engineering Discovery Environment (XSEDE), which is supported by National Science Foundation grant number ACI-1053575.}

\end{document}